\documentclass[aps,onecolumn,english,superscriptaddress,citeautoscript,preprintnumbers,amsmath,amssymb,floatfix,footinbib]{revtex4}
\usepackage{hyperref}
\hypersetup{colorlinks=true,linkcolor=red,citecolor=blue}
\usepackage{amsmath}
\usepackage{graphicx}
\usepackage{dcolumn}
\usepackage{float}

\begin{document}
\title{Magnetic field driven complex phase diagram of antiferromagnetic heavy-fermion superconductor Ce$_3$PtIn$_{11}$ }
\author{Debarchan Das}
\affiliation{Institute of Low Temperature and Structure Research, Polish Academy of Sciences, P. O. Box 1410, 50-950 Wroc{\l}aw, Poland}
\author{Daniel Gnida}
\affiliation{Institute of Low Temperature and Structure Research, Polish Academy of Sciences, P. O. Box 1410, 50-950 Wroc{\l}aw, Poland}
\author{{\L}ukasz Bochenek}
\affiliation{Institute of Low Temperature and Structure Research, Polish Academy of Sciences, P. O. Box 1410, 50-950 Wroc{\l}aw, Poland}
\author{Andriy Rudenko}
\affiliation{Institute of Low Temperature and Structure Research, Polish Academy of Sciences, P. O. Box 1410, 50-950 Wroc{\l}aw, Poland}
\author{Marek Daszkiewicz}
\affiliation{Institute of Low Temperature and Structure Research, Polish Academy of Sciences, P. O. Box 1410, 50-950 Wroc{\l}aw, Poland}
\author{Dariusz Kaczorowski}
\email{d.kaczorowski@int.pan.wroc.pl}
\affiliation{Institute of Low Temperature and Structure Research, Polish Academy of Sciences, P. O. Box 1410, 50-950 Wroc{\l}aw, Poland}


\begin{abstract}
We present the results of our comprehensive investigation on the antiferromagnetic heavy-fermion superconductor Ce$_3$PtIn$_{11}$ carried out by means of electrical transport, heat capacity and ac magnetic susceptibility measurements, performed on single-crystalline specimens down to 50 mK in external magnetic fields up to 9~T. Our experimental results elucidate a complex magnetic field -- temperature phase diagram which contains both first- and second-order field-induced magnetic transitions and highlights the emergence of field stabilized phases. Remarkably, a prominent metamagnetic transition was found to occur at low temperatures and strong magnetic fields. In turn, the results obtained in the superconducting phase of Ce$_3$PtIn$_{11}$ corroborate an unconventional nature of Cooper pairs formed by heavy quasiparticles. The compound is an almost unique example of a heavy fermion system in which superconductivity may coexist microscopically with magnetically ordered state.
\end{abstract}

\maketitle

\section{Introduction}

Investigation of anomalous physical properties at quantum critical point (QCP) in heavy fermion (HF) systems has been at the forefront of contemporary condensed matter research as it may enlighten the fascinating physics involved with these unusual observations at this precarious point of instability \cite{Gegenwart, Si, Kuchler, Steglich1}. QCP is associated with continuous quantum phase transition (QPT) which, unlike classical phase transition, is driven by tuning non-thermal parameters \cite{Coleman, Sachdev}. In this framework, Ce based HF systems are of particular interest as some of them exhibit superconductivity along with enhanced quasiparticle effective mass, known as heavy fermion superconductors (HFSC) \cite{Mathur, Steglich2, White, Steglich3, Petrovic, Bauer1, Kaczorowski}, where the pairing mechanism is believed to be mediated by magnetic fluctuations rather than  phonons as in case of conventional superconductors \cite{Monthoux}. In addition, due to unstable nature of $4f$ orbitals, magnetic ordering in these systems can be tuned by controlling parameters of non-thermal origin, like pressure, chemical doping or magnetic field, leading the system to a QCP where various unusual features such as non-Fermi liquid (NFL) behavior, unconventional superconductivity, etc., are witnessed \cite{Ohashi, Paglione, Dong,Kaluarachchi}. The majority of the vast number of Ce based HF systems reported in the literature, posses just one position for Ce ions in their crystallographic unit cell. The ground states in these compounds basically result from the interplay between two competitive interactions, namely Ruderman-Kittel-Kasuya-Yosida (RKKY) and Kondo interactions, which can qualitatively be understood in terms of the Doniach phase diagram \cite{Doniach}.  However, the scenario does not appear trivial when there are more than one Ce sites in the crystal structure. The theoretical study by Benlagra et al. on the interplay between two different Kondo effects originating from two inequivalent Kondo sublattices dictates that in different ranges of the conduction-band filling these interactions can be competitive or cooperative, which leads to a fairly complex phase diagram \cite{Benlagra}.

Investigation of novel Ce based compounds hosting multiple inequivalent Ce sites has received considerable attention from the scientific research community in recent times, as they showcase diverse unusual ground state properties at low temperatures. Different local environment of Ce ions leads to different hybridization strengths, which spark the possibility of having distinctly different ground states for each individual inequivalent Ce ion. For instance, separate AFM orderings of two Ce sublattices has been reported in Ce$_5$Ni$_6$In$_{11}$ \cite{Tang}, while Ce$_9$Ru$_4$Ga$_5$ contains three independent Ce sites where two of them order antiferromagnetically at $T_{\rm {N}}$ =3.7~K and a third one exhibits valence fluctuations \cite{dk}. Most remarkably, in the compounds Ce$_3$TIn$_{11}$ (T = Pt and Pd), bearing two inequivalent Ce sites, a coexistence of AFM order and HFSC has recently been discovered \cite{Prokle, Custers2, Kratochv}, which set a new playground for comprehensive studies on mutual relationship between magnetism and superconductivity in Ce-based HF systems.

The compound Ce$_3$PtIn$_{11}$ belongs to the homologous series of phases Ce$_n$T$_m$In$_{3n+2m}$ (T stands for $d$-electron transition metal), which encompasses a large variety of intriguing materials including CeCoIn$_5$ \cite{Petrovic}, CeRhIn$_5$ \cite{Park}, Ce$_2$PdIn$_8$ \cite{Kaczorowski, Kaczorowski2, Kaczorowski3}, Ce$_2$CoIn$_8$ \cite{Chen}. It crystallizes in a tetragonal structure with space group P$4/~mmm$, which harbors two inequivalent Ce sites \cite{Tursina}. Interestingly, Ce$_3$PtIn$_{11}$ exhibits two successive AFM phase transitions at $T_{\rm {N1}}$ = 2.2~K and $T_{\rm {N2}}$ = 2~K, followed by a superconducting transition at $T_{\rm c}$ = 0.32~K \cite{Prokle, Custers2}. It was arguably speculated that the magnetic ordering is associated with one of the Ce sublattices, while the other one is responsible for HF behavior and superconductivity \cite{Custers2}. Remarkably, under hydrostatic pressure the compound exhibits a quantum critical behavior with QCP located at a critical pressure $p_{\rm c}$ = 1.3~GPa, where $T_{\rm N}$~$\rightarrow$~0 and $T_{\rm c}$ becomes maximum, with normal state electrical resistivity showing NFL behavior \cite{Custers2}.

The spectacular low-temperature properties of Ce$_3$PtIn$_{11}$ motivated us to study the effect of applied magnetic field on the complex AFM ordering and the superconducting state. In particular, our research was aimed to explore  field effect on the two AFM transitions and construct a relevant magnetic phase diagram. Electrical transport and thermodynamic measurements were performed on single crystalline specimens at temperatures down to 50~mK. Here, we report the results of our investigations on Ce$_3$PtIn$_{11}$, which conjointly indicate a complex $H-T$ phase diagram and unconventional SC state coexisting with the AFM ordering.

\section{Results and discussions}

\subsection{Crystal structure}

We carried our a comprehensive investigation to determine the precise crystal structure of Ce$_3$PtIn$_{11}$ using an anisotropic model. Even though the crystal structure of this compound was reported previously\cite{Kratochvilova}, here we present the results of our X-ray diffraction measurements performed on a tiny crystal which offers very little absorption and high transmission factors of the X-ray beam. This allowed us to reach a high refinement quality of the observed pattern. The crystallographic data are presented in Table 1. The lattice parameters are in good agreement with the previous study\cite{Kratochvilova}. The anisotropic displacement parameters (see Table 2) indicated that the structure is well defined. The atomic positions and interatomic distances are presented in Table 2 and 3, respectively.

\begin{table}[htb!]
	\centering
	\caption [Refinement-para]{Crystal data and structure refinement for Ce$_3$PtIn$_{11}$.}
	\label{table:Cryst_struc}
	\vskip .5cm
	\addtolength{\tabcolsep}{+5pt}
	\begin{tabular}{c c c c c c c}
		\hline
		\hline
		\\
	Empirical formula &Ce$_3$PtIn$_{11}$\\
	Formula weight &1878.47\\
	Temperature &295(2) K\\
	Wavelength &0.71073 \AA\\
	Crystal system &Tetragonal\\
	Space group &P$4/~mmm$\\
	Unit cell dimensions & $a$ = 4.6977(2)~\AA, $c$ = 16.8813(13)~\AA \\
	Volume &372.54(4) \AA$^3$\\
	Z &1\\
	Density (calculated) &8.373 Mg/m$^3$\\
	Absorption coefficient &34.904 mm$^{-1}$\\
	F(000) &791\\
	Crystal size &0.066 $\times$ 0.031 $\times$ 0.020 mm$^3$\\
	Theta range for data collection &3.621 to 26.697$^{\circ}$.\\
	Index ranges &-5$\leq$h$\leq$5, -5$\leq$k$\leq$5, -21$\leq$l$\leq$21\\
	Reflections collected &10014\\
	Independent reflections &294 [R(int) = 0.0674]\\
	Completeness to theta = 25.242$^{\circ}$ &98.8\% \\
	Absorption correction & Gaussian\\
	Max. and min. transmission &0.583 and 0.290\\
	Refinement method &Full-matrix least-squares on F$^2$\\
	Data / restraints / parameters &294 / 0 / 21\\
	Goodness-of-fit on F$^2$ &1.133\\
	Final R indices [I $>$2$\sigma$(I)] & R1 = 0.0313, wR2 = 0.0717\\
	R indices (all data) &R1 = 0.0345, wR2 = 0.0732\\
	Largest diff. peak and hole &2.104 and -1.775 e.\AA$^{-3}$\\
	\hline
	\hline
	\end{tabular}
\end{table}

\begin{table}[htb!]
	\centering
	\caption [Aniso-para]{Anisotropic displacement parameters (\AA$^2\times$10$^3$), atom coordinates and isotropic displacement parameters (\AA$^2\times$10$^3$) 	in the unit cell of Ce$_3$PtIn$_{11}$.  The anisotropic displacement factor exponent takes the form: -2$\pi^2$[h$^2$a*$^2$U$^{11}$+...+2hka*b*U$^{12}$]. The isotropic displacement parameter, U(eq), is defined as one third of  the trace of the orthogonalized U$^{ij}$ tensor.}
	\label{table:Ce3PdIn11-aniso}
	\vskip .5cm
	\addtolength{\tabcolsep}{+5pt}
	\begin{tabular}{c c c c c c c c c c c c c}
		\hline
		\\
		& U$^{11}$& U$^{22}$& U$^{33}$&U$^{23}$&U$^{13}$&U$^{12}$&x&y&z&U(eq)\\
		\hline
	
	Ce(1)&8(1)&8(1)&10(1)&0&0&0&	0&	0&	0.2776(1)&9(1)\\
	Ce(2)&10(1)&10(1)&10(1)&0&0&0&	0&	0&	0&10(1)\\
	Pt(1)&12(1)&12(1)&13(1)&0&0&0&	0&	0&	0.5&12(1)\\
	In(1)&9(1)&19(1)&11(1)&0&0&0&	0.5&	0&	0.4117(1)&	13(1)\\
	In(2)&10(1)&10(1)&20(1)&0&0&0&	0.5	&0.5&	0.2778(1)&	13(1)\\
	In(3)&10(1)&18(1)&12(1)&0&0&0&	0.5&	0&	0.1379(1)&	13(1)\\
	In(4)&11(1)&11(1)&18(1)&0&0&0&	0.5&	0.5&	0&	13(1)\\
		\hline
	\end{tabular}
\end{table}

\begin{table}[htb!]
	\centering
	\caption [bond length]{Interatomic distances [\AA] for Ce$_3$PtIn$_{11}$.}
	\label{table:Ce3PdIn11-coordinates}
	\vskip .5cm
	\addtolength{\tabcolsep}{+5pt}
	\begin{tabular}{c c c c c }
		\hline
		\\
		Ce(1)-In(1)&	3.2623(12)\\
		Ce(1)-In(2)&	3.32177(14)\\
		Ce(1)-In(3)&	3.3287(12)\\
		Ce(2)-In(3)&	3.3065(9)\\
		Ce(2)-In(4)&	3.32177(14)\\
		Pt(1)-In(1)&	2.7821(7)\\
		In(1)-In(1)$^2$&	2.982(2)\\
		In(1)-In(2)&	3.2591(15)\\
		In(1)-In(1)$^1$&	3.32177(15)\\
		In(2)-In(1)$^1$&	3.2590(15)\\
		In(2)-In(3)&	3.3320(15)\\
		In(3)-In(4)&	3.3065(9)\\
		In(3)-In(3)$^1$&	3.32177(15)\\		
		\hline
		$^1$-y,x,z; $^2$-x+1,-y,-z+1
	\end{tabular}
\end{table}

\subsection{AFM ordering}

In order to have an in-depth understanding of the complex AFM ordering in Ce$_3$PtIn$_{11}$, we performed heat capacity measurements, $C(T)$, in different magnetic fields applied along the crystallographic $c$ direction in the tetragonal unit cell. Fig. 1a displays the temperature evolution of $C/T$ taken in zero field, which manifests two successive AFM transitions at $T_{\rm {N1}}$ = 2.08~K and $T_{\rm {N2}}$ = 1.93~K. It is worth noting that both critical temperatures are slightly lower than those reported by Prokle\v{s}ka et al. \cite{Prokle}. We tentatively anticipate that this discrepancy could be associated with very tiny structural features, like atomic disorder, vacancies, local distortion, which may affect the electronic state of the Ce ion (labeled in Ref.~\cite{Custers2} as Ce2) responsible for the AFM ordering in Ce$_3$PtIn$_{11}$. Another important remark that should be highlighted is the absence of any anomaly in $C/T (T)$ at 10~K (see Fig. 1a), which proves that the sample investigated was free from CeIn$_3$ impurity phase.

\begin{figure}[htb!]
	\centering
	\includegraphics[width=16cm, keepaspectratio]{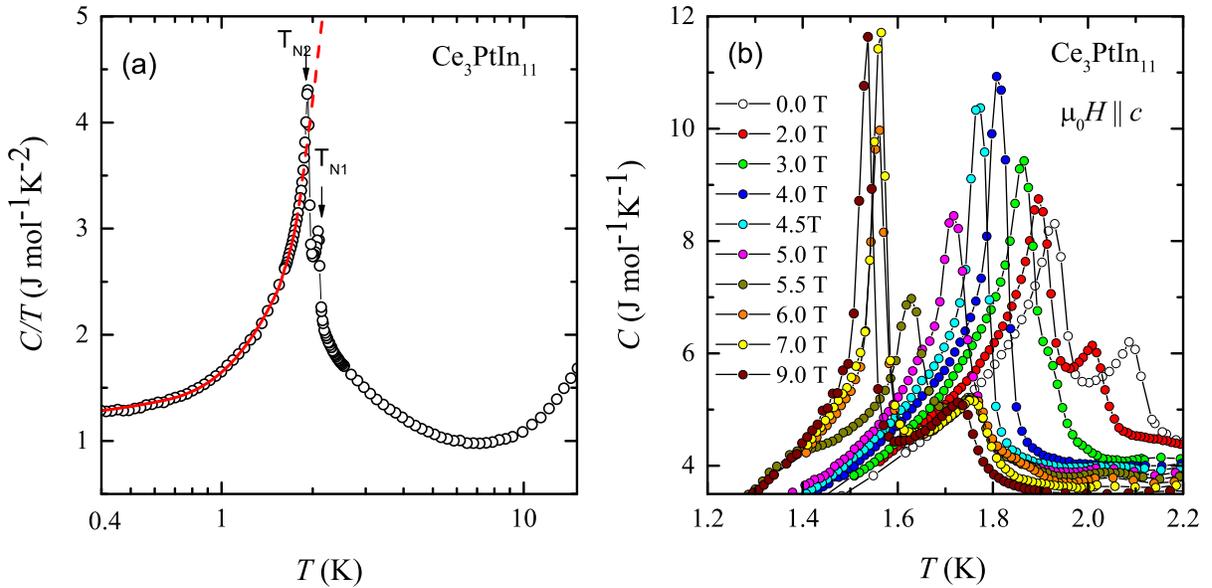}
    \caption{\label{fig:Heat Capacity}(Color online)(a) Temperature variation of the specific heat over temperature ratio measured for single-crystalline Ce$_3$PtIn$_{11}$ in zero external magnetic field. Solid red line represents least-square fittings with Eq.~1. (b) Low-temperature dependencies of the specific heat of single-crystalline Ce$_3$PtIn$_{11}$ taken in various magnetic fields applied along the crystallographic $c$ axis.}
\end{figure}

The heat capacity of antiferromagnetic Kondo lattices in the ordered state is often modeled by the formula \cite{Medeiros, Continentino}

\begin{equation}
C= \gamma_{\rm {AFM}}T + c\Delta_{\rm {SW}}^{7/2}~\sqrt{T}~\textup{exp}\left(\frac{-\Delta_{\rm {SW}}}{T}\right)
\times\left[1+\frac{39}{20}\frac{T}{\Delta_{\rm {SW}}}+\frac{51}{32}\left(\frac{T}{\Delta_{\rm {SW}}}\right)^2 \right]
\end{equation}

\noindent where the first term signifies the contribution from the heavy quasi-particles and the second term represents AFM magnons contribution with $\Delta_{\rm {SW}}$ being the gap in the spin wave spectrum. The approach assumes a spin wave dispersion relation of the form $\omega = \sqrt{\Delta_{\rm {SW}}^2+ D k^2}$, where $D$ is the spin wave stiffness. The coefficient $c$ in Eq.~1 is related to $D$ as $c~\sim~D^{-3}$~~\cite{Medeiros, Continentino}. The least-squares fitting of the above formula to the experimental data of Ce$_3$PtIn$_{11}$ measured below $T_{\rm {N2}}$ is shown in Fig.~1a by the red solid line. The so-obtained parameters are $\gamma_{\rm {AFM}}$ = 1.33(1)~J/mol~K$^2$, $c$ = 76(2)~mJ/mol~K$^4$ and $\Delta_{\rm {SW}}$ = 5.9(2)~K. The value of $\Delta_{\rm {SW}}$ is in fairly good agreement with that of estimated by Custers $et~al.$ ($\Delta_{\rm {SW}}$ = 7.35~K \cite{Custers2}). Furthermore, this gap value is comparable with the energy scale of the critical field (~ 4.8 T) of the metamagnetic transition at 0.5 K (see below), which is coming out to be ~3.6 K.

As can be inferred from Fig. 1b, with increasing magnetic field the two transitions first shift towards lower temperatures, elucidating AFM nature of the ordering. Remarkably, in a field of about 4~T, these two lambda-type features merge into a single sharp peak of a first-order character, that appears at a temperature which is  decreasing with ramping field up to 5~T. In stronger fields, the latter singularity again splits into two anomalies at temperatures which continue to decrease with increasing field. In addition to the sharp anomalies seen, we observed another very broad anomaly near 1.4~K in the specific heat data measured in an applied field of 5.5~T. This feature can be associated with the metamagnetic transition (see below).

The temperature dependence of the electrical resistivity, $\rho (T)$, of Ce$_3$PtIn$_{11}$, measured with the current flowing within the basal $ab$ plane of the tetragonal unit cell of the compound, is presented in Fig.~2a. In the interval 30-100~K, the resistivity decreases with increasing temperature in a logarithmic manner signifying the dominance of Kondo type spin-flip scattering processes. At higher temperatures, $\rho(T)$ forms a broad minimum, and susequently an increase of the resistivity with rising $T$ is seen. This feature is most likely a direct consequence of phonon contribution competing with the Kondo interactions\cite{Coqblin,Kaczorowski4}. Near 15~K, $\rho (T)$ forms a broad maximum that can be attributed to a crossover from incoherent to coherent Kondo regime that is a generic feature of Ce-based Kondo lattices, including the phases Ce$_n$T$_m$In$_{3n+2m}$. For instance, similar feature has been seen in CeCoIn$_5$ ($n$ = 1 and $m$ = 1) \cite{Petrovic}, Ce$_2$PdIn$_8$ ($n$ = 2 and $m$ = 1) \cite{Kaczorowski} and Ce$_3$PdIn$_{11}$ ($n$ = 3 and $m$ = 1) \cite{Kratochvilova}.

\begin{figure}[htb!]
	\centering
	\includegraphics[width=17cm, keepaspectratio]{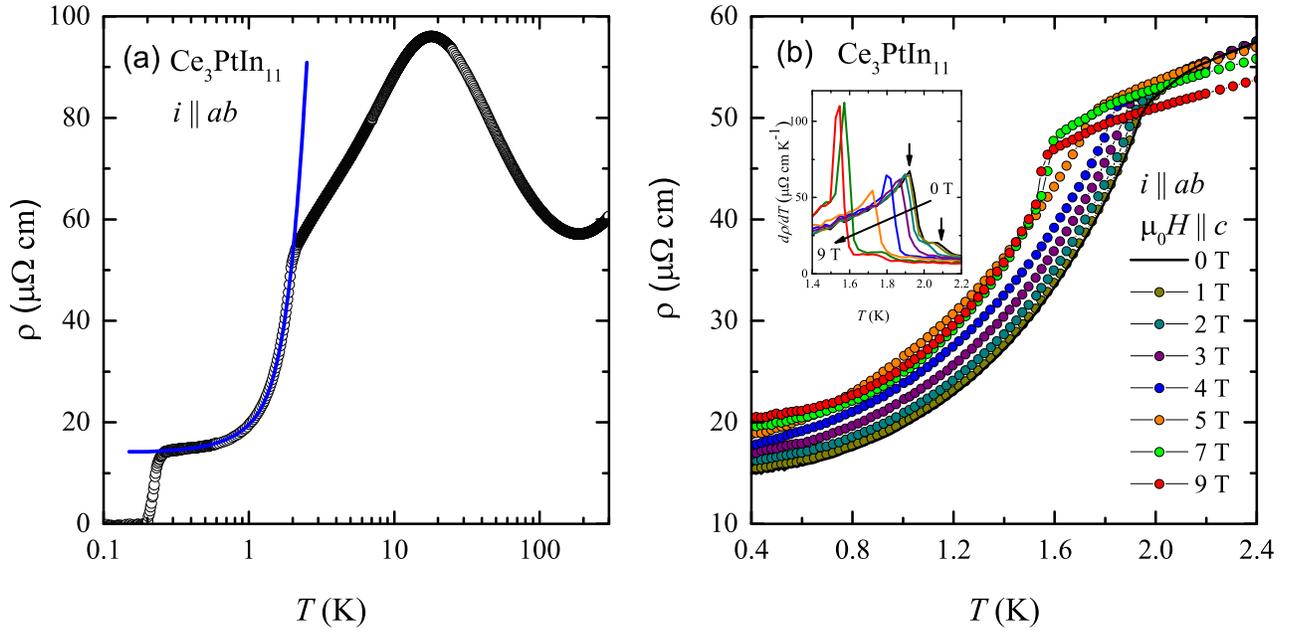}
   \caption{\label{fig:Electrical_Resistivity}(Color online) (a) Temperature dependence of the electrical resistivity of single-crystalline Ce$_3$PtIn$_{11}$ measured with electric current flowing within the tetragonal $ab$ plane. The solid blue line represents least-square fittings with Eq.~2. (b) Low-temperature electrical resistivity of Ce$_3$PtIn$_{11}$ measured within the tetragonal plane in different magnetic fields applied along the crystallographic $c$ axis. The inset represents the temperature evolution of $\frac{d\rho}{dT}$ at different fields as mentioned in the main panel. }
\end{figure}

Successively, a distinct drop in $\rho (T)$ is seen at 2.1~K caused by the reduction in the spin-disorder scattering due to the AFM ordering below $T_{\rm {N1}}$. Another sharp resistivity drop down to zero occurs at $T_{\rm c}$ = 0.23~K, and signals the onset of the superconducting state. As displayed in Fig.~2a, at temperatures $T_{\rm c} < T < T_{\rm {N1}}$, the $\rho(T)$ data can be modeled by the formula \cite{Fontes}

\begin{equation}
\rho(T) = \rho_0 + A~T^2 + b~\Delta_{\rm {SW}}^2~\sqrt{\frac{T}{\Delta_{\rm {SW}}}}~\textup{exp}\left(-\frac{\Delta_{\rm {SW}}}{T}\right)
\times~\left[1+\frac{2}{3}\frac{\Delta_{\rm {SW}}}{T}+\frac{2}{15}\left(\frac{\Delta_{\rm {SW}}}{T}\right)^2\right] ,
\label{eq:2}
\end{equation}

\noindent where the second term accounts for Fermi liquid contribution, and the third term represents scattering of conduction electrons on AFM spin-wave excitations with an energy gap $\Delta_{SW}$ in the magnon spectrum (here, phonon contribution was neglected because of very low temperature range considered). The coefficient $b$ in this expression is related to the spin-wave stiffness $D$ as $b~ \sim D^{-\frac{3}{2}}$ \cite{Fontes}. Least-squares fitting of Eq.~2 to the experimental data (see the figure) yielded: $\rho_0$ = 14.0(1) $\mu\Omega$cm, $A$ =5.4(5) $\mu\Omega$~cm~K$^{-2}$, $b$ = 9.1(3) $\mu\Omega$~cm~K$^{-2}$ and $\Delta_{\rm {SW}}$ = 10.8(1)~K. The so-obtained value of $\Delta_{\rm {SW}}$ is somewhat larger than that derived from the heat capacity data (see above). However, considering difference in methods/experiments involved, these values are fairly comparable. In turn, the quite large value of $A$ manifests the significance of electron-electron scattering in this compound, as generally expected for HF systems.

\begin{figure}[htb!]
	\centering
	\includegraphics[width=16cm, keepaspectratio]{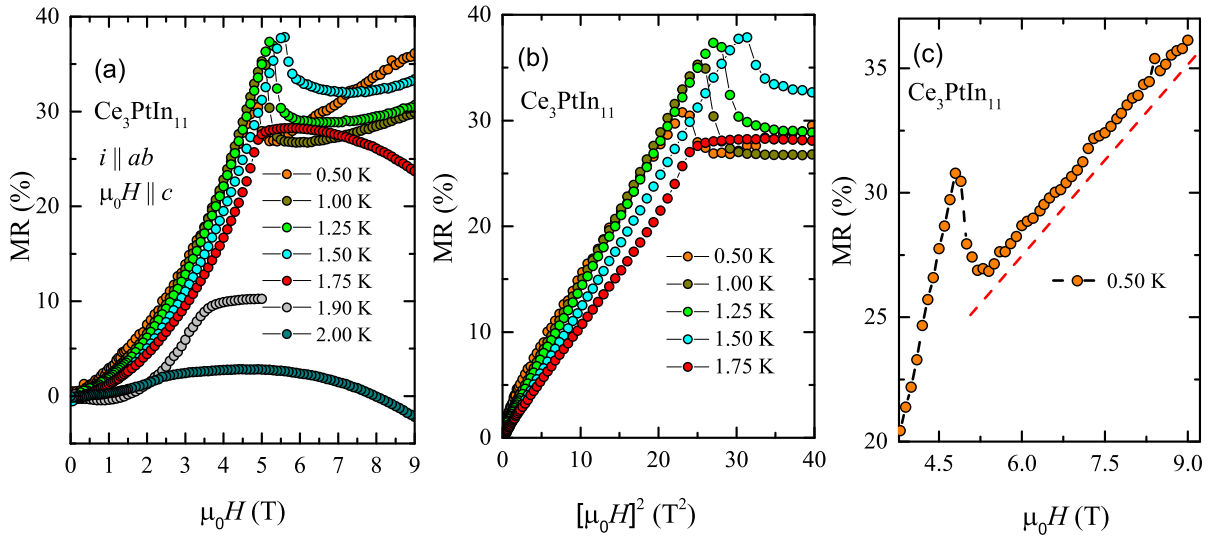}
	\caption{\label{fig:Magnetoresistance}(Color online) (a) Magnetic field dependencies of the transverse magnetoresistance of single-crystalline Ce$_3$PtIn$_{11}$ measured at several temperatures in the AFM state with electrical current flowing within the tetragonal $ab$ plane and magnetic field applied along the crystallographic $c$ axis. (b) Low-field section of the magnetoresistance isotherms from panel (a) plotted as a function of squared field. (c) High-field section of the 0.5~K magnetoresistance isotherm from panel (a). Dashed red line emphasizes a strong linear dependency observed above the metamagnetic transition.}
\end{figure}

In accordance with the heat capacity results, the AFM transition seen in $\rho (T)$ gradually shifts towards lower temperatures with increasing strength of transverse (applied along the crystallographic $c$ axis) magnetic field (see Fig.~2b). Notably, in fields 0~T~$\leq~\mu_0H~\leq$~2T and 7~T~$\leq~\mu_0H~\leq$~9~T, two distinct anomalies in d$\rho$/d$T (T)$ are observed (note the arrows in the inset to Fig.~2b), highlighting two separate phase transitions. It is worth pointing out that the field variations of the critical temperatures coincide very well with those derived from the heat capacity data. In addition, another observation which demands special note is the sudden drop in $\rho (T)$ at the ordering temperature observed in applied fields of 7~T and 9~T that gives rise to sharpening of the peak  in d$\rho$/d$T (T)$. This finding is in perfect concert with the features seen in $C(T)$ (compare Fig.~1b) suggesting a first order transition.

In order to gain further insight on the field-dependent critical behavior observed in the specific heat data, the transverse magnetoresistance (MR = $\frac{\rho(\mu_0H)-\rho(0)}{\rho(0)}$) of Ce$_3$PtIn$_{11}$ was measured with electric current flowing within the crystallographic $ab$ plane and external magnetic field applied along the four-fold axis. The results obtained in the AFM state are shown in Fig.~3a. Far below $T_{\rm {N1}}$, MR is positive and increases with ramping field in a quadratic manner (cf. Fig.~3b) up to a critical field $\mu_0H_{\rm c}$, at which a pronounced peak is observed. The positive value of MR is consistent with the AFM ordering in the system, and the MR~$\propto (\mu_0H)^2$ dependence can be attributed to the influence of magnetic field on the energy dispersion of the AFM spin waves, as predicted by Yamada and Takada within the random phase approximation \cite{Yamada}. In turn, the MR singularities at $\mu_0H_{\rm c}$ manifest metamagnetic-like phase transitions. Remarkably, the value of $\mu_0H_{\rm c}$ shows a non-monotonic temperature dependence: on rising $T$ up to 1.5~K, it increases but with further increasing temperature, it slightly decreases. Close to $T_{\rm {N2}}$, the feature in MR is quite broadened and then disappears.

\begin{figure}[htb!]
	\centering
	\includegraphics[width=16cm, keepaspectratio]{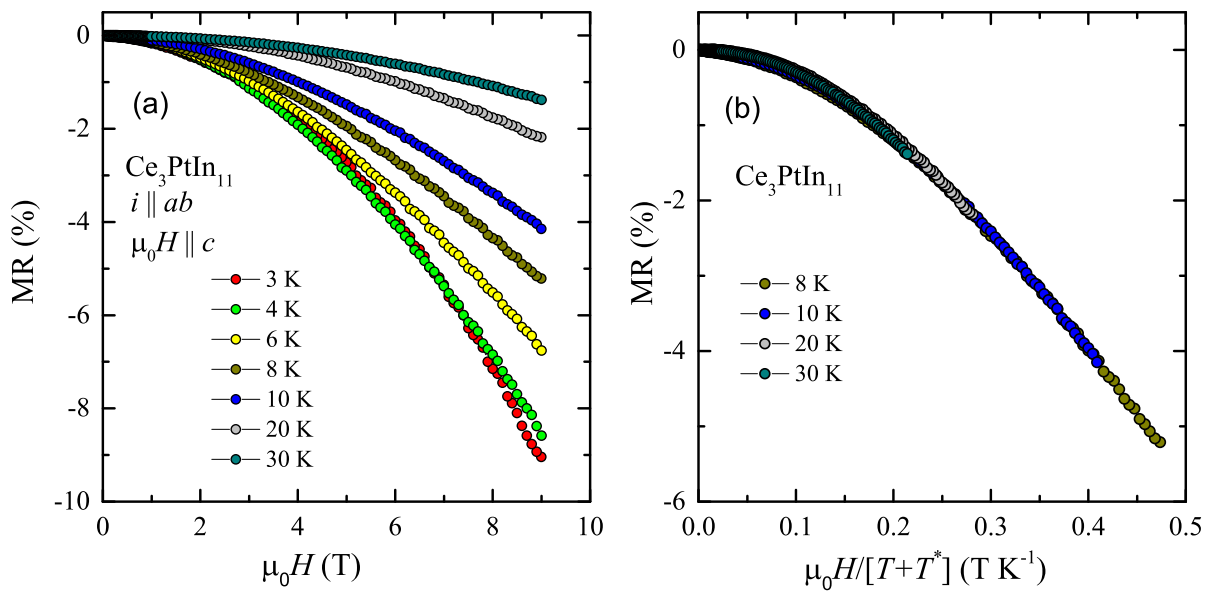}
	\caption{\label{fig:Magnetoresistance_aboveTN}(Color online) (a) Magnetic field dependencies of the transverse magnetoresistance of single-crystalline Ce$_3$PtIn$_{11}$ measured at several temperatures in the paramagnetic state with electrical current flowing within the tetragonal $ab$ plane and magnetic field applied along the crystallographic $c$ axis. (b) Schlottmann-type plot of the magnetoresistance isotherms from panel (a).}
\end{figure}

Fig.~3c shows the high-field MR data measured at temperatures 0.5~K (one can expect that in these conditions the scattering of conduction electrons on spin fluctuations is strongly damped). As can be inferred from the figure, above the metamagnetic transition, MR has a linear dependency with field.  This feature accounts for some unusual kind of cyclotron motion of the charged particles. It is worth recalling that linear MR may arise in systems with small concentrations of charge carriers having small effective masses, in regime of the electrical transport involving only the lowest Landau level \cite{Abrikosov}. Clearly, this mechanism should be ruled out for Ce$_3$PtIn$_{11}$ which is a HF compound. Another possibility for linear MR arises for gapless materials with linear energy spectrum \cite{Abrikosov2,Parish}. In the case of the compound studied, the spin wave gap $\Delta_{\rm{SW}}$ = 10.1~K was found (from electrical resistivity), and hence also the latter scenario cannot be justified. Actually, the physical origin of the linear contribution to the magnetoresistance of Ce$_3$PtIn$_{11}$, being dominant at $T$ = 0.5~K (see Fig.~2c) remains unclear. It is worthwhile mentioning that similar behavior of MR was found before for antiferromagnetic Ce$_2$PdGa$_{12}$ \cite{Gnida} but also for this compound no explanation of this unusual feature was given. Above 1.5~K, the MR isotherms change their overall shapes in strong fields.

Fig.~4a displays the transverse magnetoresistance of single-crystalline Ce$_3$PtIn$_{11}$ measured as described above at few temperatures in the paramagnetic state. At each temperature, MR is negative and its absolute value decreases with increasing temperature. Such a behavior of MR is expected for a Kondo compound due to freezing-out of the spin-flip scattering by external magnetic field. Remarkably, as shown in Fig.~4b, all the MR isotherms taken at $T \geq$ 8~K can be projected onto a single curve by plotting the MR data as a function of $\mu_0H/(T+T^*)$, where the parameter $T^*$ is the characteristic temperature, usually considered as an approximate measure of the Kondo temperature \cite{Schlottmann}. This so-called Schlottmann-type scaling was applied to Ce$_3$PtIn$_{11}$ yielding $T^*$ = 12~K.

\subsection{Magnetic phase diagram}

Fig. 5 presents the magnetic field - temperature phase diagram of Ce$_3$PtIn$_{11}$, constructed based on the results of thermodynamic and electrical transport measurements. Interestingly, the phase boundary constructed from the MR data, $i.e.$, describing metamagnetic transition (MMT), evidences a first order like phase transition which is quite remarkable. Now, in order to properly understand the field evolution of the AFM ordering temperature and emergence of new field-stabilized magnetic phases, the phase diagram can be divided into three regions as pointed out in the inset of Fig. 5. Initially, two AFM transitions occurring at $T_{\rm {N1}}$ and $T_{\rm {N2}}$ shift to lower temperatures with increasing the magnetic field strength. Then, in the field range 4~T$\leq \mu_0H\leq$5~T these two transitions merge into a single feature at $T_{\rm M}$ further  decreasing with ramping field. This finding is quite consistent with the expectation that with increasing field, Zeeman energy is increased and when it exceeds the energy of the intersite-coupling strength, the long-range ordering is turned into a field-induced ferromagnetic state. However, with magnetic field above 5~T, one observes another peculiarity, namely the latter transition again splits into two well separated anomalies seen at $T_{\rm {M1}}$ and $T_{\rm {M2}}$. The positions of these singularities systematically decrease with raising field, at least up to 9~T. Notably, the height of the peak at $T_{\rm M1}$ systematically decreases with increasing field, while the peak at $T_{\rm {M2}}$ rapidly sharpens on going from $\mu_0H =$ 5.5~T to $\mu_0H \geq$ 6~T. Furthermore, at high fields, the shape of the latter anomaly, observed both in the specific heat (see Fig. 1b) and the electrical resistivity (see Fig. 2b), resembles to a first order like transition. This may hint towards possible rearrangement of the Fermi surface sparking the possibility of a field induced Lifshitz transition in this system. Further detailed investigations probing the Fermi surface geometry such as quantum oscillations or angle-resolved photoemmision spectroscopy are needed to verify that conjecture.

\begin{figure}[htb!]
	\centering
	\includegraphics[width=10cm, keepaspectratio]{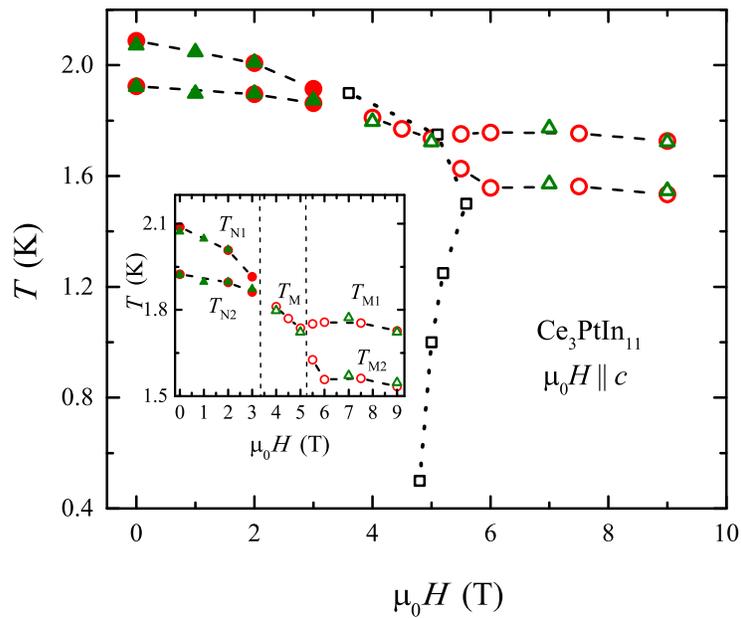}
	\caption{\label{fig:Phase diagram}(Color online) Magnetic phase diagram of Ce$_3$PtIn$_{11}$ constructed from the heat capacity (red circles), electrical resistivity (green triangles) and magnetoresistance (black squares) data. Full and open symbols are used to distinguish between the second-order and first-order transitions, respectively. The inset represents a magnified part of the main panel without MMT. Vertical dotted lines serve as a guide for the eye, illustrating three distinct regions discussed in the text.}
\end{figure}

Remarkably, the $H-T$ phase diagram constructed for Ce$_3$PtIn$_{11}$ bears striking similarities with the magnetic phase diagrams of intensively studied HF antiferromagnets CeRhIn$_5$ and Ce$_2$RhIn$_8$\cite{Cornelius}. In particular, a common feature is the presence of both first- and second-order field-induced magnetic transitions. Therefore, the phase diagram of Ce$_3$PtIn$_{11}$ turns out to be quite remarkable as it may suggest the existence of competing order parameters in this material. In order to explore this peerless feature, further investigation involving neutron diffraction and muon spin relaxation/reorientation will be essential. Another direction for future studies would be investigating the thermodynamic and transport properties of Ce$_3$PtIn$_{11}$ in other magnetic field orientations, in order to shed some light on the expected anisotropy of the field stabilized magnetic phases emerging in this material.

\subsection{Superconducting phase}

\begin{figure}[htb!]
	\centering
	\includegraphics[width=8.5cm, keepaspectratio]{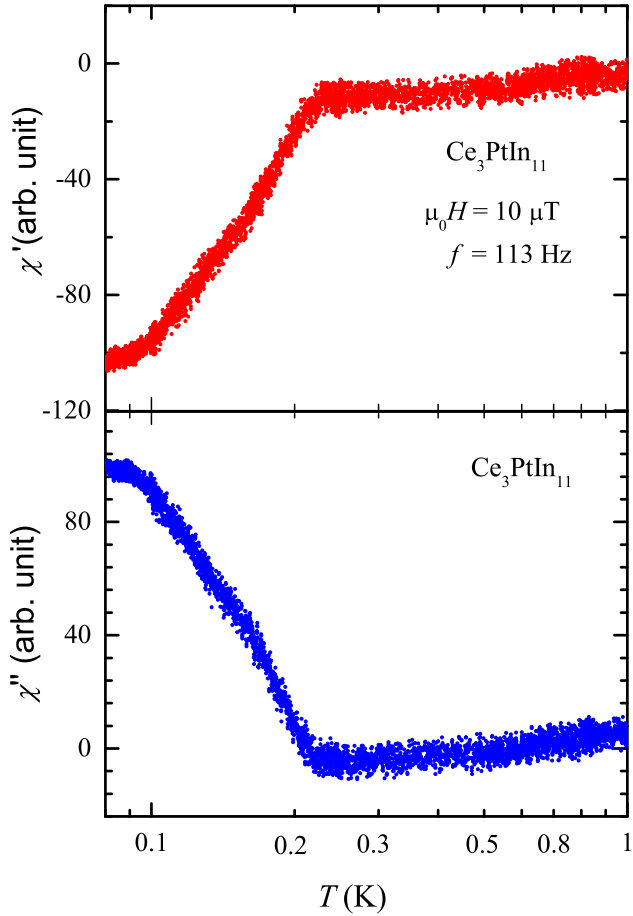}
	\caption{\label{fig:AC}(Color online) Temperature variations of the real ($\chi ^{\prime} $) and imaginary ($\chi ^{\prime\prime} $) components of the ac magnetic susceptibility of single-crystalline Ce$_3$PtIn$_{11}$ measured in an excitation field of 10~$\mu$~T and frequency of 113 Hz.}
\end{figure}

Fig.~6 depicts the real and imaginary components of the dynamic magnetic susceptibility of Ce$_3$PtIn$_{11}$ measured in zero steady magnetic field with an ac field of 10~$\mu$~T oscillating with frequency of 113~Hz. Both a clear diamagnetic signal in $\chi ^{\prime} (T)$ and a pronounced upturn in $\chi ^{\prime\prime} (T)$ below $T_{\rm c}$ = 0.23(1)~K manifest the onset of bulk superconductivity in the specimen measured. Remarkably, the so-derived value of $T_{\rm c}$ is in perfect agreement with the electrical resistivity data described below, however distinctly smaller than $T_{\rm c}$ = 0.32~K reported in the literature \cite{Prokle, Custers2}. Possible source of this discrepancy may be related to some tiny structural features, as suggested in the previous section in the context of the magnetic critical temperatures. Clarification of this intriguing issue requires comprehensive crystallographic investigations of single crystals of Ce$_3$PtIn$_{11}$ grown in different batches.

\begin{figure}[htb!]
	\centering
	\includegraphics[width=16cm, keepaspectratio]{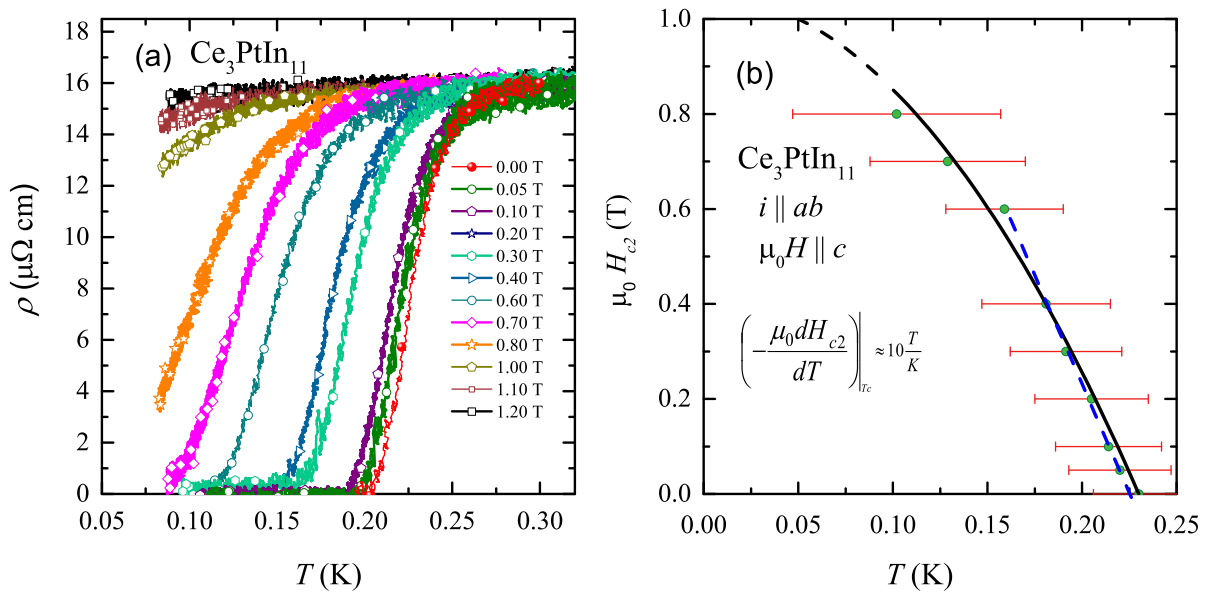}
	\caption{\label{fig:SC_phase}(Color online) (a) Low-temperature dependencies of the electrical resistivity of single-crystalline Ce$_3$PtIn$_{11}$ measured with electric current flowing within the tetragonal $ab$ plane in different external magnetic fields applied along the $c$ axis. (b) Temperature variation of the upper critical field in single-crystalline Ce$_3$PtIn$_{11}$ for the electric current and magnetic field geometry as indicated in panel (a). The black solid line represents the fitting of the observed data with a mean field type expression as described in the text. The blue dashed line manifests the initial slope.}
\end{figure}

Fig.~7a shows the ultra-low temperature dependence of the electrical resistivity of Ce$_3$PtIn$_{11}$ measured with electric current flowing in the tetragonal $ab$ plane and external magnetic field applied along the crystallographic $c$ axis. Clearly, with increasing field strength the superconducting transition gradually broadens and shifts to lower temperatures. The critical temperature, defined at the midpoint of the drop in $\rho (T)$, is equal to $T_{\rm c}$ = 0.23(2)~K, which is in good agreement with the ac magnetic susceptibility data. Plotting the change of $T_{\rm c}$ in the magnetic field one can derive the temperature variation of the upper critical field in the specimen studied. From the results presented in Fig.~7b, the initial slope of the $\mu_0 H_{\rm {c2}} (T)$ dependence near $\mu_0 H = 0$ is found out to be $\mu_0\left(\frac{\delta H_{\rm {c2}}}{\delta T} \right)_{T=T_{\rm c}}$ $\approx$ -10~T/K. The overall $\mu_0H_{\rm {c2}}(T)$ dependence can be well approximated using a mean-field type expression

\begin{equation}
\mu_0 H_{\rm {c2}}(T)= \mu_0 H_{\rm {c2}}(0)\left[1-\left(\frac{T}{T_{\rm c}}\right)^2\right]
\label{eq:Meanfield}
\end{equation}

yielding $\mu_0 H_{\rm {c2}}(0)$ $\approx$ 1.1~T which is much larger than the Pauli-Cologston-Chandrasekhar limiting field $\mu_0~H_{\rm{P}}$ = $1.86T_{\rm c}$ = 0.43~T.

Now from the formula

\begin{equation}
\xi_{\rm {GL}} = \left[\phi_0/2\pi\mu_0 H_{\rm {c2}}(0)\right]^{1/2}  ,
\label{eq:GL}
\end{equation}

\noindent where $\phi_0=h/2e$ is the flux quantum, one can estimate Ginzburg-Landau (GL) coherence length in Ce$_3$PtIn$_{11}$ to be $\xi_{\rm {GL}}$ = 88~$\textup{\AA}$. 

\begin{table}[htb!]
	\centering
	\caption [Supercon para]{Comparison of the superconducting parameters of Ce$_3$PtIn$_{11}$ (this work), Ce$_2$PdIn$_8$ \cite{Kaczorowski} and CePt$_3$Si \cite{Bauer2}}
	\label{table:Ce3PdIn11-superconducting}
	\vskip .5cm
	\addtolength{\tabcolsep}{+5pt}
	\begin{tabular}{c c c c c c c}
		\hline
		Parameters & Ce$_3$PtIn$_{11}$ & Ce$_2$PdIn$_8$ & CePt$_3$Si  \\
		\hline
		\\
		$T_{\rm{c}}$ (K) & 0.23 & 0.68 & 0.75\\
		$\gamma$ (J/mol~K$^2$) & 1.33 & 1.0 & 0.335  \\
		$\mu_0 H_{\rm{c2}}(0)$ (T) &1.1 & 4.8 & 4.0 \\
		$-\mu_0\frac{dH_{\rm{c2}}(0)}{dT}$ (T/K) & 10.0 & 14.3 & 8.5  \\
		$\xi_{\rm{GL}}$ ($\textup{\AA}$) &88 & 82 & 81 \\
		\hline
		\hline
	\end{tabular}
\end{table}

The key characteristics of the superconducting state in Ce$_3$PtIn$_{11}$ are gathered in Table 4, where they are compared with those reported for Ce$_2$PdIn$_8$ \cite{Kaczorowski} and CePt$_3$Si \cite{Bauer2}. A close resemblance of these various superconducting parameters with those of the well established HF superconductors hints towards a possible unconventional origin of the superconductivity that emerges in Ce$_3$PtIn$_{11}$ even within the AFM ordered state.

\section{Conclusions}

In summary, the results of our detailed investigation of AFM ordering and superconducting phase in single-crystalline Ce$_3$PtIn$_{11}$ elucidate a likely coexistence of AFM and superconductivity in this compound. The specific heat and electrical transport data collected in various applied magnetic fields conjointly establish a complex magnetic phase diagram with several distinct field stabilized magnetic phases. It sparks the possibility of finding competing order parameters near the critical field in this compound. This observation is quite remarkable considering the uniqueness of such phase diagram in the existing literature. Furthermore, the magnetotransport data collected at low temperatures revealed  a metamagnetic transition followed by a linear field dependence. This feature implies an unusual kind of cyclotron motion. 

The electrical resistivity and ac magnetic susceptibility data obtained in the superconducting state in Ce$_3$PtIn$_{11}$ indicated unconventional superconductivity with the key parameters similar to those reported for well-established heavy-fermion superconductors. Thus, Ce$_3$PtIn$_{11}$ turns out to be one of the rare examples of HF systems where superconductivity coexists with bulk magnetic order. Further detailed investigations involving microscopic techniques such as muon spin rotation, neutron diffraction and photoemission spectroscopy are called for in order to address the unusual and unique features witnessed in Ce$_3$PtIn$_{11}$.

\section{Methods}

Single crystals of Ce$_3$PtIn$_{11}$ were grown using In flux, as outlined by Kratochv$\acute{\textup{i}}$lov$\acute{\textup{a}}$ $et~al$. \cite{Kratochvilova}. The crystals selected for physical properties measurements were examined by x-ray diffraction (XRD) an Oxford Diffraction four-circle single crystal diffractometer equipped with a CCD detector using graphite-monochromatized Mo-K$\alpha$ radiation. The raw data were treated with the CrysAlis Data Reduction Program (version 1.171.38.34a). The intensities of the reflection were corrected for Lorentz and polarization effects. The crystal structures were solved by direct methods and refined by full-matrix least-squares method using SHELXL-2014 program\cite{Sheldrick}. The atoms were refined using anisotropic displacement parameters. Their chemical composition was checked by energy-dispersive X-ray (EDX) analysis using a FEI scanning electron microscope equipped with an EDAX PV9800 microanalyzer. The XRD and EDX results confirmed the expected stoichiometry and the crystal structure of the compound, in line with the literature data \cite{Prokle, Custers2}.

The electrical resistivity was measured over the temperature interval 0.4 to 300 K and in magnetic fields up to 9 T using a standard ac four-probe technique implemented in a Quantum Design PPMS platform. In order to probe the superconducting state, a Cryogenic Ltd. $^3$He-$^4$He dilution refrigerator was employed to carry out electrical resistivity measurements down to 50~mK in applied fields up to 1.2~T.  Furthermore, the ac magnetic susceptibility was measured in the same dilution fridge. Heat capacity measurements were performed in the range 0.35-20 K in fields up to 9~T using relaxation method and the PPMS equipment.

\section{Acknowledgment}
This work was supported by the National Science Centre (Poland) under research grant No. 2015/19/B/ST3/03158.

{\bf Author contributions statement}\\
D.K. conceived the experiments and supervised the research. D.D., D.G., {\L}.B. and A.R. performed the physical properties measurements and contributed to the analysis of the experimental data. M.D. conducted the structural characterization of the crystals. D.D. and D.K. wrote the manuscript with notable input from all the other authors.\\

{\bf Competing financial interests:} The authors declare that they have no competing interests.

\end{document}